# Using math in physics: 3. *Anchor equations*

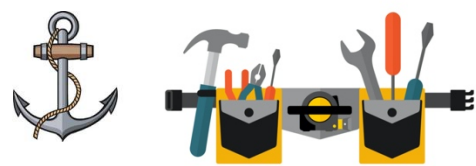

*Edward F. Redish,*
University of Maryland - emeritus, College Park, MD

An important step in learning to use math in science is learning to see physics equations as not just calculational tools, but as ways of expressing fundamental relationships among physical quantities, of coding conceptual information, and of organizing physics knowledge structures. In this paper I discuss the role of basic *anchor equations* in introductory physics and show some examples of how to help students learn to use them.

Anchor equations provide stable starting points for thinking about whole blocks of physics content. For example, Newton's 2nd law is an anchor equation. It is the central principle that provides a *foothold* -- a starting point for organizing our discussion of classical mechanics at the introductory level. Some of the conceptual ideas that are coded in this anchor are shown in Fig. 1.

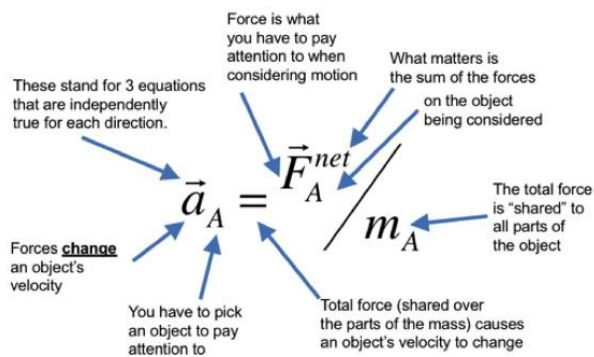

*Figure 1. Some conceptual knowledge coded in Newton's 2nd law.*

This paper is part of a series on the topic of learning to use mathematics as a thinking tool in science. The first paper (paper 0) gives an overview of the student difficulties involved and the basic tools for analysis and instruction.[1] In papers 1 and 2, I discuss *ontological* issues - having to do with what kind of a thing our symbols in a physics equation stand for and how they are assigned numerical values — dimensional analysis,[2] and estimation.[3] These papers focus on helping students establish an intuition for the physical meaning of measurement and scales.

In this paper and the next, I focus on *epistemological* issues - issues about the nature of knowledge in physics and how it's structured and used to generate new knowledge. In this paper, I discuss how conceptual knowledge about the structure and relationship of physical measurements are coded into fundamental equations. In the next, I discuss how equations are used as a part of the construction of simplified mental models that provide a starting point for thinking about physical systems.[4]

## Students rarely see the central role equations play in the structure of physics knowledge

Building a good understanding of science involves learning different kinds of knowledge:

- Facts
- Procedures
- Mechanism
- Synthesis

Facts and procedures are important, but they're only the first steps in beginning to build scientific skills. Unfortunately, much of school and introductory science classes focus solely on these elements. Since they are well suited to memorization, students often come away with the epistemological misconception that memorization is all there is to science.

*Mechanism* — building causal stories and extended chains of reasoning where each step sets up the conditions for the next — is rarely a part of introductory science classes. While introductory biology classes teach a number of multi-step processes, they tend to be learned more as a memorized string of processes rather than what we would describe in physics as a mechanism of chained causality, where each step produces the conditions that lead to the next.

*Synthesis* — putting lots of knowledge together in a structured and organized way that *makes sense* — is the heart and soul of building scientific knowledge and learning to use it. Students accustomed to memorizing lots of terms in introductory biology or lots of molecular structures and reactions in introductory chemistry may miss the coherence in physics, both among its different parts and with everyday experience.

Introductory physics can be of special value to students in other disciplines because we can introduce mechanistic thinking and synthesis in simpler situations than can biology, chemistry, earth science, or even engineering. We can show that mechanistic model building and synthesis are often tied together by powerful equations, so students can learn the value of mathematics, not as something to be memorized, but as something to support complex reasoning, analysis, and powerful descriptions of physical systems.

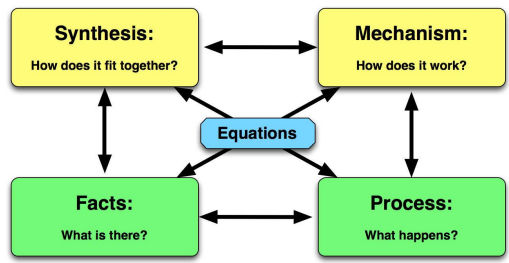

*Figure 2. A graphical representation of knowledge structure in science. In physics, symbolic math and equations play a coordinating role.*

In physics, a few critical equations can help synthesize a lot of knowledge about a content topic. I refer to these fundamental equations as ***anchor equations*** to emphasize their special role as stable fixed points in organizing knowledge. When I call on an anchor equation in slides or in our text (the NEXUS/Physics wiki[5]), I put the anchor icon shown at the top of this article to help students realize how valuable those equations are and how often we refer back to them.

Identifying anchor equations and learning how to use them can help students make the transition to a more thoughtful use of math in science. They have four specific benefits:

- Anchor equations encourage synthesis and principle-based reasoning.
- Anchor equations help students learn to blend physical concepts with math symbols.
- Anchor equations code conceptual knowledge.
- Anchor equations organize problem-solving knowledge.

## Anchor equations encourage synthesis and principle-based reasoning

Students in introductory physics rarely have experience in powerful principle-based reasoning, and many see physics (and all the sciences) as an incoherent and hard-to-memorize collection of facts and highly localized procedures. While both biology and chemistry have powerful organizing principles (for example, evolution in biology and chemical bonding rules in chemistry) and occasionally have problems that use some math with them, their fundamental statements are qualitative rather than quantitative. Students often have a hard time seeing equations as also providing principles to reason with.

In physics, our powerful theoretical frameworks are tied to mathematics and fundamental equations — Newton's 2nd law, the work-energy theorem, Maxwell's equations, the Schrödinger equation, Einstein's equations of general relativity. This kind of mathematics-based organizational knowledge structure is not common in introductory classes in the other sciences. If we want to help students learn synthesizing skills and principle-based reasoning, we have to be explicit about it. Identifying anchor equations and encouraging students to start with them (instead of jumping to a memorized special case equation) can help.

## Anchor equations help students learn to blend physical concepts with math symbols

We want students to see equations as a way of saying something physical, not just as a way to calculate. Often, textbooks present all of the equations that are useful for calculation in their end-of-chapter summaries. This sends students the message (an "epistemological meta-message") that these many equations are all independent calculational tools. It encourages memorization and discourages them from making the challenging transition to thinking synthetically with math.

### Example: Conceptual kinematic equations

One example of this is the kinematic equations. End-of-chapter summaries often include lots of equations for calculating velocities and distances under multiple conditions. An alternative approach is to avoid the calculationally useful kinematic equations and instead use only the conceptually more natural kinematic definition equations. Students can do all the same calculations, but they start with equations that make conceptual sense, map their knowledge of the physical system into the equation, and only then manipulate the math to get their answer. This can help students learn to build the physics/math blend and build their confidence that they can reason with math, rather than simply calculate with it.

The basic kinematics concepts are

- In a given time interval the position may change (velocity)
- In a given time interval the velocity may change (acceleration).

The direct mapping of these physical concepts into math symbology are given by the word equations:

1. The average velocity is given by the change in position (How far did you move?) divided by the time interval (How long did it take to do it?).
2. The average acceleration is given by the change in velocity (How much did it change?) divided by the time interval (How long did it take to do it?).

For motion along a line (1D), we write these symbolically as follows

1. $<v> = \frac{\Delta x}{\Delta t} \quad v = \frac{dx}{dt}$
2. $<a> = \frac{\Delta v}{\Delta t} \quad a = \frac{dv}{dt}$

where we use the notations "<...>" to indicate the average over an extended time interval and "Δ" to indicate the change in the following quantity. My class has calculus as a prerequisite so I include the definition of instantaneous velocities.[6]

Once students learn these coding rules,[7] these equations come to be seen as natural and sensible definitions and, if a student has managed to construct the blend, these equations will not need to be memorized. If forgotten, they can be easily reconstructed from the conceptual definition. This illustrates the difference between memorizing an equation and "knowing" it, something the students may have little previous experience with.

Two additional equations allow one to generate all the usual kinematic equations: the definition of change and the evaluation of average in the case of constant change.

The definition of the delta can be seen as making physical sense:

The final value = the initial value + the change in the value

or, translated into an equation

$x_f = x_i + \Delta x$ or easily rearranged into $\Delta x = x_f - x_i$

While this looks trivial, it is only so if you are mapping meaning to the symbols. Students who are still learning to do this often write down something else (perhaps with a + instead of a - or in the wrong direction or they ignore the delta entirely or even put in a ½ from the confusion with the average). This happens much more often than you might expect.

The second equation students need is the average when the change is at a uniform rate (easily shown geometrically):

The average value = ½ (the initial value + the final value)
(if the quantity is changing at a constant rate)

or, translated into an equation

$<v> = 1/2\,(v_i + v_f)$

Students also get this wrong, replacing it by the difference extraordinarily often. When they do this, it's a clear signal they are relying on memorization and not sense-making (not making the blend). The fact that this is an epistemological error ("What kind of knowledge should I use here?") rather than a knowledge error can easily be demonstrated by asking them first, "What's the average of 10 and 12?" and then, "If the velocity is constant, what should the average be?"

Students often hold on very strongly to their incorrect equation, often not having the epistemological resource, "If it doesn't make sense, the equation I used must be wrong." Rather, they rely on, "I can't be expected to make sense of an equation, so somehow my sense-making intuition must be wrong." **Developing the resource that equations and intuitions must match is one of the most important skills we can help students develop.** Sometimes you do have to modify your intuitions, but sometimes you've remembered or used the equation incorrectly.

To help students learn to use equations by blending physics and math, rather than just picking an equation with the right variables[8] and plugging in, we need to give problems that are not easily solved by pulling up a single standard equation. I give some examples in the instructional materials below. More are in the supplementary materials in EPAPS.

## Anchor equations
## code for conceptual knowledge

As trained physicists who blend physics and math and can unpack the knowledge coded in an equation, we often use a shorthand that hides the content being called upon and makes an equation look like a computational tool to be memorized.

Newton's 2nd law is a great example of this. It's the fundamental powerful principle (anchor equation) that underlies all of our understanding of classical motion at scales from the molecular to the galactic, at speeds up to a significant fraction of the speed of light. It's THE principle that organizes all our knowledge of motion. And yet…

Before I realized all this, I had a tendency to just write "F = ma" for Newton's second law. I sat up and took notice, however, when, in one of my physics-for-life-sciences classes, after my discussion of springs, a student asked, "Professor, what's the difference between F = kx and F = ma?" All of a sudden a lot of what felt like weird student behavior fell into place. Why would they use F = ma for each of the different forces in a problem? Why would they tell me acceleration caused forces? And how would they ever forget it? Well, of course one reason is they were not making the blend, just using each equation as a calculational tool. But a second reason was that they were failing to identify one of the two equations as a core principle (and the other as a crude toy model).

That's when I began to write text called, "Reading the content in …" in which I identified a few anchor equations and wrote a reading for each in which I specified a basic equation and codified the conceptual physics that was coded in the equation.[9] One example is the conceptual kinematic equations discussed above. A second is Newton's 2nd law.

It's important, at least for a while, to stress the general validity of Newton 2 and to help students realize that is not just an equation to calculate the relation among two variables and a parameter, but that the full form, with diacritical markers (vectors, subscripts, … that students often ignore), specifies a lot of conceptual ideas.

One way to display this is shown in Fig. 1 at the beginning of this article.[10] Each of the detailed ways we choose to write the equation in that figure specifies a bit of conceptual knowledge that students often fail to bring to bear in solving a problem. It's not that they don't know these things, they just tend not to call on them. Reminding themselves of the full form of the law can help them recall just what aspects of the physical phenomena they need to bring to bear in building their mental blend when using this equation (it's about an object, it's about the sum of forces, directions matter, …).

To let them know that I'm serious about using equations this way, I do not accept "$F = ma$" as a correct form of N2 (they have to put "net" on the force), and I strongly encourage them to write "$a = F^{net}/m$" to remind themselves that it's forces (the sum of them) that cause acceleration and not the other way around.[11]

## Anchor equations organize problem-solving knowledge.

Many students study for each week's quiz as a way of keeping up in the class. (Exactly what I want them to do!) But many are frustrated because they feel they've done all they can — reviewing the previous week's work, creating long lists of everything we covered — but still wind up doing poorly on the Monday quiz. They come to me for advice on how to prepare. Then I have them where I want them — unsatisfied with memorizing and ready to try something new!

From both of the earlier examples (the conceptual kinematic equations and the detailed form of Newton 2), we can see that these equations code conceptual knowledge. But they can also help organize the basic ideas that students have to learn to use the equations in practical problem-solving situations. This approach also helps them in doing their homework when they are not sure how to get started.

If a student approaches a problem by trying to call on the long list of "the things we've learned that I have to memorize" at the end of a typical kinematics chapter, they can get confused and lost, especially since the equations can look contradictory if they don't pay attention to the conditions attached to each.[12] This is especially true if the problem involves multiple objects, time intervals, distances, and velocities.

A more productive approach is to begin with the basic, conceptually reasonable definitions shown in figure 3. These equations can guide the approach to a problem, by unpacking them with the supplementary conceptual knowledge shown in the boxes. Each item in figure 3 is conceptually straightforward: a one-step translation of a sensible concept into symbol.

Students can then look at each specific object in a complex problem and decide what the initial and final values of each variable are, when they are known and when they are unknown, and set up the equations needed to solve the problem.

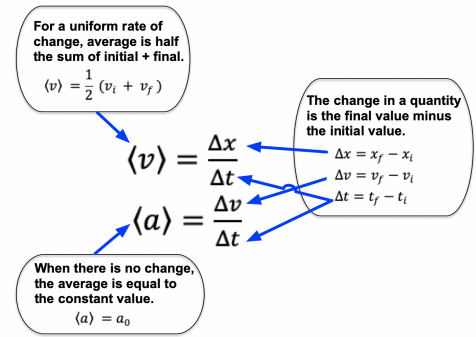

*Figure 3. Unpacking useful knowledge from the conceptual kinematic equations.*

This not only helps them solve problems with multiple objects, but encourages them to "think in the blend", identifying physical quantities with symbols. (Try this with the race car problem below.) Not only is it productive in getting to an answer, the process helps them "build the blend", associating physical quantities with mathematical symbols.

When students learn to not just review, to not just make pages and pages of review notes, but rather to take their reviews or class notes and organize them around a few core ideas — often around an anchor equation — they take a major step in transforming the way that they view the knowledge in the class. Their performance on both quizzes and exams improves dramatically, and they develop a new and surprising sense that "there's not actually that much to learn", even though there is — but it seems less if it can be easily unpacked from a small set of "things to know".

## Using anchor equations in class

To get students to value anchor equations, it's important to offer problems that are easy to set up using an anchor equation and a few straightforward manipulations rather than ones that are easy to simply put numbers into a memorized equation and calculate.

Here's an example. This problem is set up so that each step is easily solved using the kinematic anchor equations in conceptual form for average velocity and acceleration.

### A homework problem

Jeff and Dale are racing their stock cars to try to win the Daytona 500. Jeff makes a pit stop to gas up and change his tires. When he gets back on the track, he discovers that he is right next to Dale. Both are traveling at speeds of 150 mi/hr, but Dale is one full lap (2.5 miles) ahead.*

Jeff wants to catch Dale so he decides to try to accelerate slowly and uniformly (constant a) until he catches Dale -- so that after Dale has done 10 laps, Jeff will have done 11. (Jeff knows from studying physics that uniform acceleration will use less gas.) Give your answers to three significant figures.

(a) How much time does Jeff have to do it, assuming that Dale keeps a constant speed during those 10 laps?

(b) What distance will Jeff have to cover in that time?

(c) Again, assuming that Dale keeps a constant speed, what does Jeff's average speed have to be in order to catch Dale after Dale has gone 10 laps?

(d) What would Jeff's speed be when he catches Dale?

(e) What was Jeff's acceleration during the time Dale was going 10 laps?

* For those not familiar with NASCAR racing: In this race, the cars travel around a 2.5 mi oval track 200 times. During the long time that this race takes, cars have to make pit stops for service -- gas, new tires, etc.

Since one of my learning goals is "meta" — to help students learn to understand the character of physics knowledge and not just "do it" — I give an essay question on most homework sets and exams in which students are asked to think about what's going on and discuss it. Some of the problems that focus students on anchor equations are of this character.

## A homework problem

Coulomb's law describes the force between two charges. We have written it as follows:

$$\vec{F}^E_{Q \to q} = \frac{k_C qQ}{r^2_{Qq}} \hat{r}_{Q \to q} \quad (1)$$

In various textbooks you will see this written in other ways: for example

$$F = k\frac{Q_1 Q_2}{r^2} \quad (2)$$

$$F_{1 \text{ on } 2} = \frac{k|q_1||q_2|}{r^2} \quad (3)$$

A. For the expression (1), explain what each of the terms in the equation means and what they tell you about the electric force between two charges.

B. Expressions (2) and (3) differ in various ways from expression (1). Explain some of these differences (at least one for each) saying what information is represented differently (or not represented) compared to (1). What do the vertical bars mean in expression (3)?

C. Electric forces satisfy Newton's third law. Explain how this information is coded into expression (1).

D. If we analyze the force between a dipole and a charge, we discover that the force between a single charge and a dipole doesn't behave like $1/r^2$, but falls off like $1/r^3$ at long distances. Does this mean that Coulomb's law doesn't hold? When can we use Coulomb's law and when can we not use it.

More examples are given in the Supplementary Materials associated with this paper in EPAPS.

## Instructional resources

Many of the ideas for this series of paper were developed in the context of studying physics learning in a class for life-science majors. A number of problems and activities using dimensional analysis are offered in the supplementary materials to this paper. A more extensive collection of readings and activities from this project on the topic of anchor equations is available at the *Living Physics Portal*, search "Making Meaning with Mathematics: Anchor equations."

## Acknowledgements

I would like to thank the members of the UMd PERG over the last two decades for discussion on these issues. I am grateful to Khala Marshall-Watkins for suggesting the term "anchor equation." The work has been supported in part by a grant from the Howard Hughes Medical Institute and NSF grants 1504366 and 16244

---

[1] E. Redish, Using math in physics - Overview

[2] E. Redish, Using math in physics - 1. Dimensional analysis, preprint

[3] E. Redish, Using math in physics - 2. Estimation, preprint

[4] E. Redish, Using math in physics - 4. Toy models, preprint

[5] The NEXUS/Physics wiki.
https://www.compadre.org/nexusph/

[6] I don't use much more than the conceptual idea of derivative and integral. In my opinion, these are a necessary condition for learning Newtonian mechanics that can be taught in class and that doesn't require a calculus class. I explain the "d's in the derivative as "just being Δ's but when the intervals are so small we're not going to bother looking inside them."

[7] These coding rules are not trivial and students tend to ignore them at first. In particular, they tend to confuse a quantity, the change in a quantity, and the rate of change of a quantity throughout the class. This is an important distinction and it

needs to be worked on. It is discussed in detail in one of the later papers in this series.

[8] Or using the same symbols even if they aren't the right physical variables. Seeing students put in the value for a volume where what's expected is a velocity is another clear signal they are not making the blend of physics and math.

[9] These are available in the NEXUS/Physics wiki on ComPADRE. They are listed on the page describing Anchor Equations and in the Supplementary Materials on EPAPS.

[10] To see another, see the page "Reading the content in Newton's 2nd law" in the NEXUS/Physics wiki.

[11] Although "F=ma" and "a=F/m" are mathematically equivalent, they tend to activate different mental constructs. Since in math we always write "f(x) = …" students often read an equation to mean that the expression on the right is creating the quantity on the left.

[12] See the Long Homework problem in the supplementary materials on EPAPS.